\documentclass[%
unsortedaddress,
preprint,
showpacs,preprintnumbers,
 amsmath,amssymb,
 aps,
]{revtex4-1}

\usepackage{graphicx}
\usepackage{dcolumn}
\usepackage{bm}

\begin{document}


\title{Electric field control of magnetism by reversible surface reduction and oxidation reactions}

\author{K. Leistner$^{1,2}$}
 \email{k.leistner@ifw-dresden.de}
\author{J. Wunderwald$^{1,2}$}%
\author{N. Lange$^{1,2}$}%
\author{S. Oswald$^{1}$}%
\author{M. Richter$^{1}$}%
\author{H. Zhang$^{1,3}$}%
\author{L. Schultz$^{1,2}$}%
\author{S. F\"ahler$^{1,4}$}%

\affiliation{%
 $^1$IFW Dresden, P.O. Box 270116, 01171 Dresden, Germany\\
 $^2$TU Dresden, Faculty of Mechanical Engineering, Institute for Materials Science, 01062 Dresden, Germany\\
 $^3$Rutgers State University, Department of Physics and Astronomy, Piscataway, NJ 08854 USA\\
 $^4$Institute of Physics, Chemnitz University of Technology, D-09107 Chemnitz, Germany
}%


\date{\today}

\begin{abstract}
Electric control of magnetism is a vision which drives intense research on magnetic semiconductors and multiferroics. Recently, also ultrathin metallic films were reported to show magnetoelectric effects at room temperature. Here we demonstrate much stronger effects by exploiting reduction/oxidation reactions in a naturally grown oxide layer exchange coupled to an underlying ferromagnet. For the exemplarily studied FePt/iron oxide composite in an electrolyte, a large and reversible change of magnetization and anisotropy is obtained. The principle can be transferred to various metal/oxide combinations. It represents a novel approach towards multifunctionality. \begin{description}
\item[PACS numbers]75.85.+t, 75.70.-i, 75.30.Gw, 81.65.Mq, 82.45.Fk

\end{description}
\end{abstract}

\maketitle



\section{INTRODUCTION}

Though the interdependence of magnetism and electricity had already been described by Maxwell in 1861, it took more than a century until it was realized that the coupling between magnetic and electric polarization in solid matter may exponentiate the application potential of polarized materials. To date there are two main material classes exhibiting magnetoelectric effects: magnetic semiconductors \cite{Chiba} and insulating multiferroics \cite{Lottermoser}. Both utilize the weak coupling of their intrinsic bulk polarizations. This constraint leaves application above room temperature an open challenge. Two-phase multiferroics consisting of ferromagnetic/ferroelectric composites \cite{Zheng} overcome this limitation. Since (ferro-)elastic coupling mediates between both phases, they, however, cannot be applied in microsystems where thin composite films are clamped by thick, rigid substrates. Thus, alternative approaches are sought to address the growing demand on multifunctional nanosystems. For example, benefit can be drawn from the increased surface to volume ratio at the nanoscale. This idea expands the range of applicable materials towards metals with a thickness of just a few nanometers. In the bulk of metals, the free electrons shield any electric field. This does not hold for the surface atomic layers. Accordingly, electric fields can be used to change the mechanical properties of metallic nanoparticles \cite{Weissmüller,Jin}, to control magnetocrystalline anisotropy, coercivity and Curie temperature in ultrathin films \cite{Weisheit,Maruyama,Chiba2011,Bonell} and to induce surface phase changes \cite{Gerhard}. In combination with the spin torque effect \cite{Wang} or the precession of magnetization \cite{Shiota}, the electric field dependence of the magnetocrystalline anisotropy even allows bistable magnetization switching \cite{Tsymbal}. Huge effects are obtained when Curie temperature coincides with application temperature \cite{Weisheit,Maruyama,Chiba2011,Bonell}.

Here, we investigate ultrathin ferromagnetic metallic films with the advantage of a high Curie temperature. This avoids unfavorable cross-correlations originating from the high temperature dependency of magnetocrystalline anisotropy close to Curie temperature. First we show how a critical point can be adjusted in a metallic layer, which allows toggling between two magnetic states; second we demonstrate how the electric field can be used to reversibly reduce and oxidize the film and, thus, to switch the magnetic state. The electric field is applied by the established approach of electrochemical charging \cite{Weissmüller,Weisheit}. This allows understanding changes in both, magnetic anisotropy and magnetization by means of electrochemical concepts.

\section{EXPERIMENTAL}

Continuous 2 nm FePt thin films were grown by pulsed laser deposition on heated MgO(001) substrates with 3 nm Cr and 50 nm Pt buffer layers. L1$_0$ ordering results in a tetragonal distortion of the unit cell, hence this buffer can be used to align the magnetically easy c-axis out-of-plane \cite{Weisheit}. The conducting thick buffer layers allowed for the use of the films as electrodes in an electrochemical cell. Cr and Pt were deposited at a substrate temperature of 300$^\circ$C. The FePt films were deposited from a FePt alloy target at 450$^\circ$C. During post annealing the films remained in the substrate heater and temperature was held constant for times between 0 and 15 min. 

For ultrathin films, conventional magnetometry is hampered by the low overall magnetization and is impossible for a sample in an electrochemical cell. Therefore, we selected anomalous Hall effect measurements instead. This method is sensitive enough for ultrathin films and compatible with the electrochemical charging. In Hall geometry, the magnetic field was applied perpendicular to the film plane. The Hall resistivity $R_{Hall}$ for ferromagnets is the sum of ordinary ($R_0$) and anomalous ($R_{AH} = R_SM(H)$) terms: $R_{Hall} = R_0H + R_{AH}$. Hystereses presented here are corrected for the normal Hall effect, obtained by the linear increase at high magnetic fields. A current between 10 and 100 mA was applied along one in-plane direction and the Hall voltage was measured along the other in-plane direction. Thus, the magnetization component perpendicular to the film plane was probed. The Hall measurements proved to be reliable, as identical hysteresis loops were obtained by measurements of the magnetooptical Kerr effect in the polar configuration \cite{Leistnerelacta}. The Hall press contacts are expected to penetrate a surface oxide layer. The current is expected to flow through the metallic FePt and the underlying non-magnetic Pt buffer layer. The anomalous Hall effect then probes the total perpendicular magnetization component of the corresponding conducting sample region. 

For electrochemical charging, a nonaqueous electrolyte composed of 0.1-1~M~LiPF$_6$ (alternatively 0.1 M~LiClO$_4$) in dimethyl carbonate (DMC)/ ethylene carbonate (EC) (1:1) was used. The charge per surface atom was calculated from the current-voltage curves in a three electrode cell and assuming an FePt(001) surface. For in-situ Hall measurements, an electrochemical cell compatible with a Physical Properties Measurement System (PPMS) system was constructed  \cite{Leistnerelacta}. As counter electrode, Li connected to a Cu wire was used, and the electrolyte was filled in a compartment sealed with O-rings. Electrical contacts were realized by press contacts on the sample outside the electrolyte compartment. The in-situ Hall cell was assembled in an Ar-Box to avoid air contamination of the electrolyte. Films were charged between 2 and 3 V vs. Li/Li$^+$. Please note that the applied voltages here are expressed vs. a Li/Li$^+$ reference electrode, in contrast to the work of Weisheit \cite{Weisheit} using a Pt reference. The upper and lower potential limits were determined in preceding electrochemical experiments \cite{Leistnerelacta} in order to avoid Fe dissolution and electrolyte side reactions.

The chemical states of Pt and Fe were investigated by means of X-ray photoelectron spectroscopy (XPS). The measurements were carried out at a PHI 5600 CI (Physical Electronics) spectrometer which is equipped with a hemispherical analyzer operated with a typical pass energy of 29 eV and an analysis area of approx. 800 $\mu$m diameter. Monochromatic Al-K$\alpha$ excitation (350 W) was used. Peak positions for Fe$^{2+}$ and Fe$^{3+}$ have been assigned according to Ref. \cite{Yamashita}. The XPS spectra have been background corrected and normalized.

\section{ADJUSTING A CRITICAL SITUATION IN FEPT THIN FILMS}

In order to prepare a critical situation between two magnetic states, we used the competition between two independent contributions to magnetic anisotropy. The high aspect ratio of film thickness vs. lateral extension results in a large shape anisotropy preferring a magnetization direction within the film plane to reduce magnetic stray field energy. As an antagonist, the intrinsic magnetocrystalline anisotropy was used which favors magnetization along specific crystallographic directions. In order to align these crystallographic directions, we took advantage of the strong influence of the interface by epitaxial growth. We tuned the relation between shape and magnetocrystalline anisotropy in 2 nm thick FePt films by varying the annealing time $t_A$ (Figure 1). At a given annealing temperature,  $t_A$ determines the degree of order, which is essential for the strength of magnetocrystalline anisotropy in this system \cite{Kurth}. 

The $t_A$-dependent transition from an in-plane to an out-of-plane magnetization can be followed in magnetization measurements based on the anomalous Hall effect (Figure 1). As the magnetic field was oriented perpendicular to the film plane, for films with in-plane magnetization, increasing field results in a continuous rotation of magnetization out-of-plane. This is visible in an almost linear increase of the Hall signal in Figure 1 for films without annealing ($t_A$ = 0 min). In this case, saturation is reached at 1.4 T, which is precisely the anisotropy field expected for a disordered A1 FePt thin film with a saturation magnetization of 1.4 T and negligible magnetocrystalline anisotropy. Applying post annealing leads to steeper curves showing reduced in-plane total anisotropy. This can be explained by a continuous increase of magnetocrystalline anisotropy due to increasing L1$_0$ chemical order. For $t_A$ = 15 min, the Hall signal finally reveals a step-like switching, reflecting a magnetization process along an easy axis. In the present films coercivity values are low compared to granular L1$_0$ films, as in our continuous films reverse domains can easily switch the whole film once they are nucleated at a defect \cite{shima}. Saturation is thus achieved in low fields by domain wall motion. 

\begin{figure}
\includegraphics[width=0.5\columnwidth]{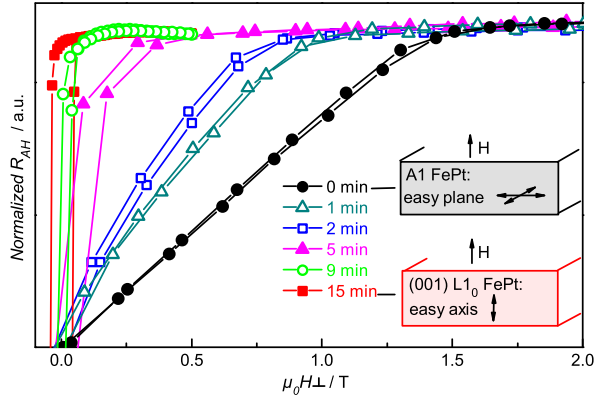}
\caption{Adjusting the critical point between shape anisotropy and magnetocrystalline anisotropy. The graph shows the first quadrant of magnetization curves measured by the anomalous Hall effect with field perpendicular to the substrate. By increasing the annealing times $t_A$ from 0 to 15 min at 450$^\circ$C the chemical order of 2 nm thick FePt films gradually changes from A1 to L1$_0$, as concluded from the transition from in-plane to out-of-plane easy axis (sketched in the inset).}

\label{ipop}
\end{figure}
 
In order to quantify the transition from in-plane ($t_A$ = 0~min) to out-of-plane ($t_A$ = 15~min) anisotropy one can use the area between magnetization curve, magnetization axis and saturation magnetization (inset in Figure 2b). This area defines an effective anisotropy constant  $K_{\rm eff}$. 

\section{CHANGES OF ANISOTROPY AND MOMENT BY CHARGING}

To obtain a large reversible variation of  $K_{\rm eff}$ by an electric field we used films annealed for an intermediate time ($t_A$ = 5 min) and charged them in an electrolyte. In contrast to previous work of Weisheit et al. \cite{Weisheit} who charged FePt in a Na/propylene carbonate (PC) electrolyte under ambient atmosphere, we used a nonaqueous Li-based electrolyte salt solved in DMC/EC and placed the complete assembly in an argon box. As described in detail in \cite{Leistnerelacta} this method allowed to expand the potential window between Fe dissolution (3 V) and electrolyte side reactions (2 V) to 1 V (in comparison to 0.6 V in \cite{Weisheit}). The magnetic properties were then measured in situ in dependence of the applied potential. After immersion in the electrolyte and without the application of an external potential the films exhibit an open circuit potential of around 3 V \cite{Leistnerelacta}. The subsequent stepwise application of more negative potentials down to 2 V corresponds to negative charging (addition of electrons).

\begin{figure}
\includegraphics[width=0.5\columnwidth]{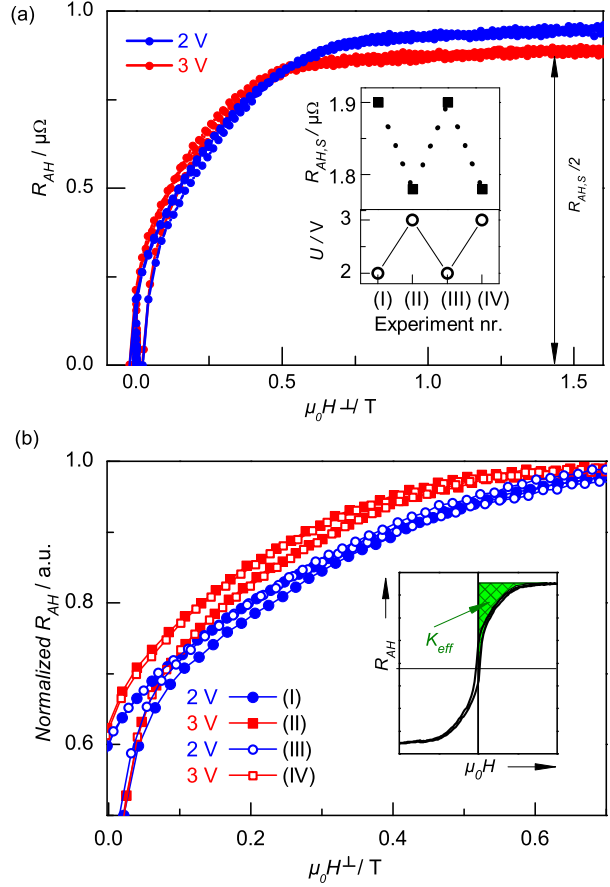}
\caption{Reversible change of magnetism by an electric field. a) The first quadrant of magnetization curves as probed by the anomalous Hall effect is shown for an FePt film ($t_A$ = 5 min) at applied potentials of 2 and 3~V in 0.1 M LiPF$_6$ in DMC/EC (1:1). The inset demonstrates reversibility of the effect by comparing the maximum anomalous Hall resistance $R_{\rm AH,S}$ as a measure for the saturation magnetization obtained in a sequence of experiments (I to IV). 
b) Normalized curves revealing the reversibility of the curvature change. The inset illustrates the applied definition of the effective anisotropy $K_{\rm eff}$.} 

\label{Fig2}
\end{figure}

The magnetic behavior of a typical film charged between 2 and 3 V is depicted in Figure 2. We observe a strong dependence of the maximum anomalous Hall resistance $R_{\rm AH,S}$, which is taken as a measure of the saturation magnetization, on the applied potential (Figure 2a). Repeated charging experiments (I to IV in the inset of Figure 2a) proof that this change in saturation magnetization is reversible. At the same time, a reversible change in the shape of the magnetization curve is observed (Figure 2b). At 3 V, the curve is steeper and reaches saturation at lower magnetic fields than at 2 V. This finding indicates a decrease of out-of-plane magnetocrystalline anisotropy for negative charging. 

To quantify the effect, we consider $K_{\rm eff}$, normalize it to its value measured at 3 V and take $\Delta K$: $\Delta K = 1 - K_{\rm eff}/K_{\rm eff}(3\;\rm V)$ as a measure for the change of the perpendicular anisotropy. The potential dependences of $\Delta K$ and $\Delta R_{\rm AH,S}=R_{\rm AH,S}/{R_{\rm AH,S}}-1$ are summarized in Figure 3. Anticipating further analysis, our data are referred to as FePt/Fe-O. If the potential is reduced from 3 to 2 V, $\Delta K$ decreases by 25\% while the saturation magnetization increases by 4\%.
\begin{figure}
\includegraphics[width=0.5\columnwidth]{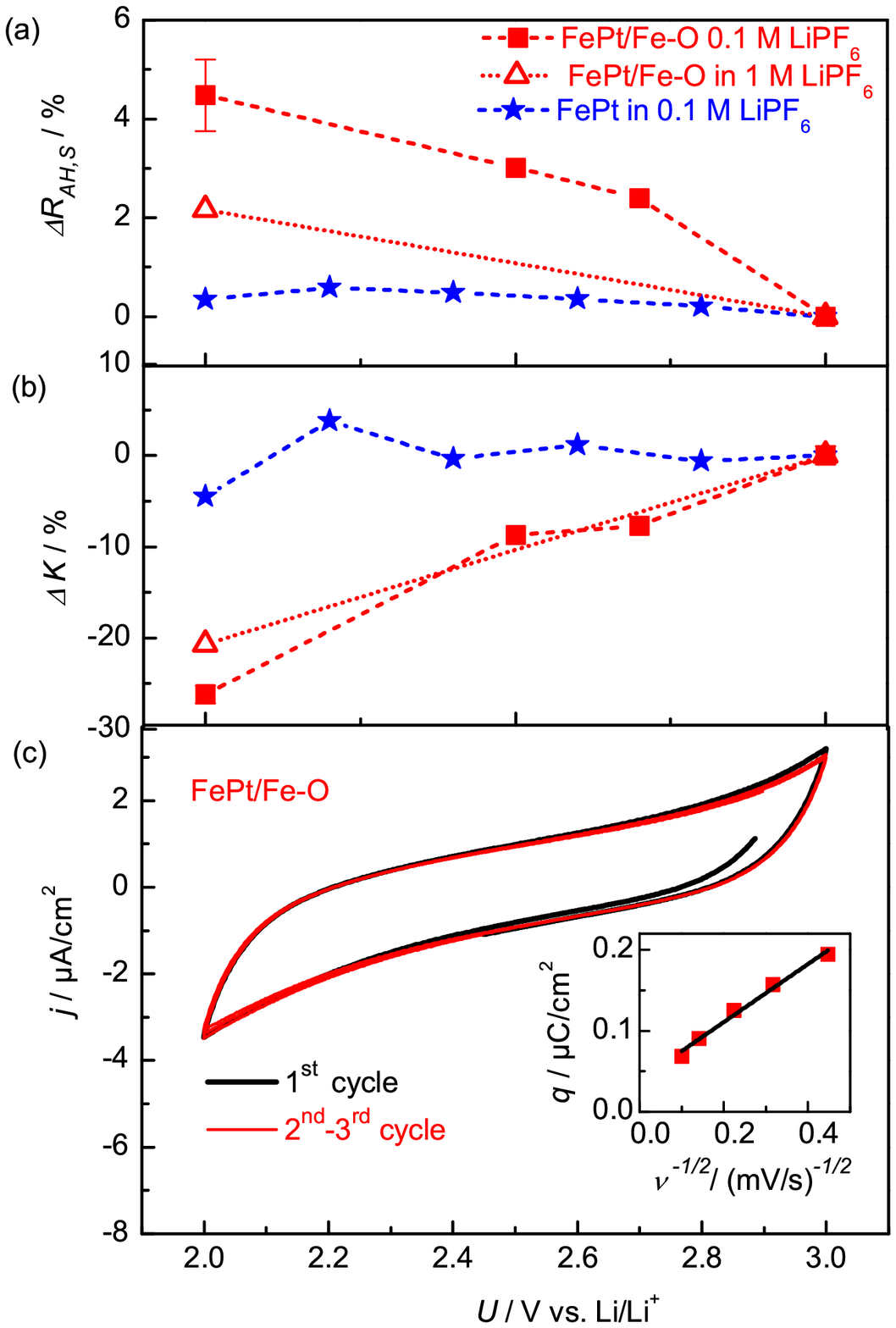}
\caption{Influence of electric field on magnetism for a composite FePt/Fe-O and for a single phase FePt sample charged in 0.1 M LiPF$_6$ in DMC/EC(1:1). For the FePt/FeO composite sample, data for 1 M LiPF$_6$ in DMC/EC(1:1) is shown in addition. a) Change of $R_{\rm AH,S}$ as a measure for the saturation magnetization. A representative error bar is shown.
b) $\Delta K$ versus applied voltage. The error lies around 1 \%. Lines are added as guide to the eye.
c) Corresponding current-voltage curve of the FePt/Fe-O composite film (scan rate 5 mV/s) and voltammetric charge $q$ versus reciprocal square root of the scan rate $\nu$.}
\label{Fig3}
\end{figure}

Follow-up experiments demonstrate that this charging effect on the magnetic properties is not restricted to one electrolyte composition. For an increased concentration of 1 M of the LiPF$_6$ salt, a qualitatively similar behaviour during charging is evident in Figure 3. Even though the effect is smaller in this case ( 2 \% change of $R_{\rm AH,S}$ and 20 \% change of $\Delta K$), a clear increase of saturation magnetization and decrease of perpendicular anisotropy is observed for negative charging. Figure 4 shows that the effect is measured also for a different electrolyte salt. As alternative to the LiPF$_6$ electrolyte salt, LiClO$_4$ has been chosen here. LiClO$_4$ offers the advantage that HF impurities, which are inevitably present in the LiPF$_6$ electrolyte and which may cause etching problems \cite{Leistnerelacta}, are avoided. Comparable to the LiPF$_6$ electrolyte, magnetization clearly increases, whereas the perpendicular anisotropy decreases for negative charging. Quantitatively the effect is weaker than in LiPF$_6$, which will be discussed within the model described in section V. 

\begin{figure}
\includegraphics[width=0.5\columnwidth]{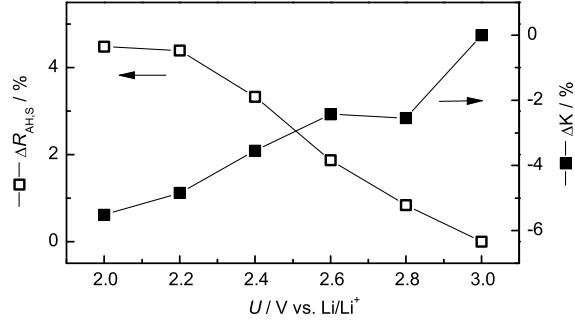}
\caption{Influence of electric field on magnetism for a composite FePt/Fe-O sample charged in 0.1 M LiClO$_4$ in DMC/EC (1:1):  Change of $R_{\rm AH,S}$ as a measure for the saturation magnetization (left axis) and $\Delta K$ (right axis) versus applied voltage.}
\label{Fig3}
\end{figure}

\section{CONTROL OF MAGNETIC EXCHANGE COUPLING \protect\\ BY REDOX REACTIONS}

At this point we can state an astonishingly strong electric-field dependence of magnetic properties, achieved by tuning ultrathin metallic films to a point of almost compensating anisotropies. We find a relative change of the anisotropy which is one order of magnitude larger than in previous experiments \cite{Weisheit}. In addition, we observe a variation of saturation magnetization similar to results on CoPd films and nanoparticles \cite{Gosh,Zhernenkov}.

As an attempt to explain the observed 4\% increase of magnetization when adding about 0.2 e$^-$ per surface atom, we performed density-functional calculations using the procedure described in \cite{Zhang}. These calculations describe the effect of band filling by the additional surface charge. They yield a marginal decrease of magnetization (-0.25\% for 0.1 e$^-$ addition) instead of a substantial increase. This inconsistency indicates that filling of the FePt bands cannot be the essential mechanism behind the measured effect.

In the following, we present experimental evidence that here the electric field control of metallic magnetism is achieved by reversible reduction/oxidation of a film with naturally oxidized surface. We show that this mechanism can explain the variation of both quantities, anisotropy and magnetization. For this aim, we investigated the surface condition of our films in more detail. As in previous experiments \cite{Weisheit,Zhernenkov}, the films had been handled under ambient conditions between growth and measurements. Though FePt is commonly considered as inert due to the noble metal component, exposure to air may result in the formation of a thin native iron oxide. Indeed, XPS measurements (Figure 5a) reveal a partial but significant shift of the Fe 2p$_{3/2}$ peak toward the position expected for iron oxides. A native iron oxide usually consists of a mixture of several iron oxide phases. The multiplet splitting, the close Fe 2p$_{3/2}$ peak positions of Fe$^{2+}$ and Fe$^{3+}$ (709.5-710.2 eV and 711.0-711.2 eV, respectively \cite{Yamashita,Roosendaal}) in combination with the low signal do not allow to accurately distinguish the kind and amount of the iron oxide phases from this XPS spectrum. From controlled oxidation of thin Fe films it is known, however, that Fe$^{2+}$ dominates close to the Fe interface and Fe$^{3+}$ forms closer to the oxide surface \cite{Roosendaal}. Accordingly, we refer to these samples in a general way as FePt/Fe-O composites. As a cross check, we investigated samples directly transferred into the Ar-box to avoid oxidation. As no peak shift towards iron oxides is observed in XPS in this case (Figure 5, data set b)), we denote these samples single phase FePt. When repeating the electric field experiments for a single phase FePt sample, only minor variations of magnetization and anisotropy are observed (Figure 3a and b). Hence, the large variations observed for the FePt/Fe-O samples indeed have to be attributed to the presence of iron oxide and not to an electronic band filling of the metallic FePt.

\begin{figure}
\includegraphics[width=0.5\columnwidth]{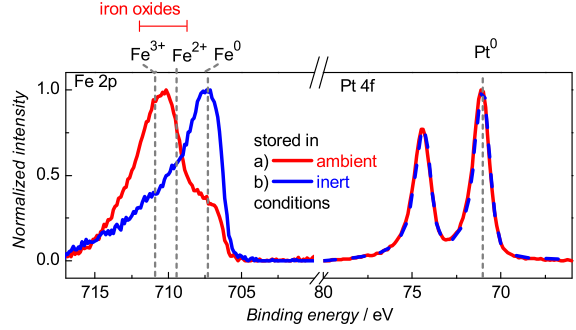}
\caption{Fe 2p$_{3/2}$ and Pt 4f$_{5/2,7/2}$ XPS spectra for 2 nm FePt films stored under a) ambient conditions and b) Ar atmosphere.
}

\label{XPS}
\end{figure}

To reveal the role of the oxide layer during charging, comparative XPS measurements have been carried out after deposition, after sole contact to the electrolyte, and after charging to 2 V (Figure 6). The oxide component of the as deposited state (Figure 6, data set a)) is smaller here than in Figure 5 (data set a)), which indicates a thinner native oxide layer thickness for this sample. Immersion in the electrolyte without external potential (at the open circuit potential, Figure 6b) does not change the Fe 2p$_{3/2}$ peak position and an almost identical spectrum is obtained. This means that neither the iron oxide layer nor the metallic FePt is altered by a chemical reaction in the electrolyte. It becomes obvious however, that charging at 2~V leads to a significant shift of the Fe 2p$_{3/2}$ peak from the peak positions of iron oxides in the direction of metallic Fe. The chemical state of Fe is thus clearly altered by the external potential, which proves the presence of an electrochemical reaction. A more detailed understanding of this electrochemical reaction and its relation to the magnetic property changes is obtained in the following discussion of electrochemical concepts in combination with the results of cyclovoltammetry, XPS and magnetic measurements.

Iron oxide exhibits a rich electrochemistry involving redox reactions between the different iron oxides, hydroxides, and metallic iron. The formation of specific iron species in an electrolyte is related to the applied potential. In aqueous solution, reduction of iron oxides and hydroxides is known to occur in the following order for decreasing potential \cite{Jacintho2007}: FeO$_x$(OH)$_{3-2x}$, FeOOH, and Fe$_2$O$_3$ $\rightarrow$ Fe$_3$O$_4$ $\rightarrow$ metallic Fe. The oxidized species FeO$_x$(OH)$_{3-2x}$, FeOOH, and Fe$_2$O$_3$ are non-magnetic or weakly ferromagnetic. At decreased potentials, these Fe$^{3+}$ species are reduced to ferrimagnetic Fe$_3$O$_4$ (which contains Fe$^{3+}$ and Fe$^{2+}$) with a bulk saturation polarization $J_S$ of 0.7 T. At even lower potential a further reduction to metallic and thus ferromagnetic Fe ($J_S = 2.1$ T in the bulk) takes place. For surface-near regions these reduction processes can be reversible and subsequent anodic charging again oxidizes the surface to Fe$_3$O$_4$ and FeOOH \cite{Jacintho2007}. Such a potential induced reduction from a non-magnetic Fe$^{3+}$ iron oxide surface layer to a ferromagnetic Fe surface layer could well explain the observed reversible increase of saturation magnetization and the shift in the XPS spectra for negative charging to 2 V.

\begin{figure}
\includegraphics[width=0.5\columnwidth]{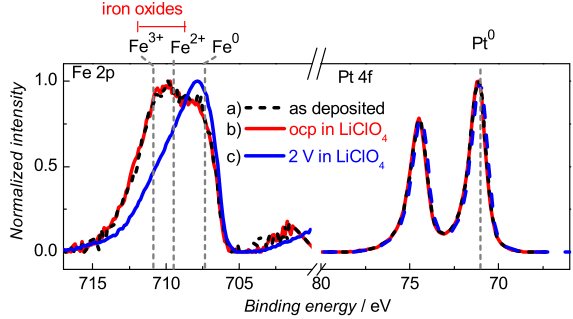}
\caption{Fe 2p$_{3/2}$ and Pt 4f$_{5/2,7/2}$ XPS spectra for 2 nm FePt films after immersion in 0.1 M LiClO$_4$ in DMC/EC in comparison to the a) as prepared state. b) is obtained after 2 h immersion without external potential (= open circuit potential, ocp) and c) after 2 h immersion at a constant potential of 2 V.      }
\label{Fig3}
\end{figure}
In aqueous solution, protons are involved in the reduction reaction (e.g. Fe$_3$O$_4$ + 8 H$^+$ + 8 e$^-$ $\leftrightarrow$ 3 Fe$^0$ + 4 H$_2$O)\cite{Jacintho2007}. In the present nonaqueous Li-based electrolyte protons are not available, but instead Li$^+$ ions can take their part. This mechanism was e. g. reported for the reduction/oxidation of transition metal nanoparticles in 1 M LiPF$_6$ in DMC/EC \cite{Poizot2000}. There, a lithium-driven reversible oxidation/reduction reaction has been described for cobalt oxide according to: CoO + 2 Li$^+$ + 2 e$^-$ $\leftrightarrow$ Li$_2$O + Co. We propose that a similar oxidation/reduction of iron oxide is responsible for the observed reversible change of magnetization. It should be noted that Li incorporation is unlikely to play a role here, as it starts below 2 V in Fe$_2$O$_3$ \cite{Pernet} and Fe$_3$O$_4$ \cite{Lim}, and as it would be accompanied by a severe decrease (instead of the observed increase) of magnetic moment \cite{Sivakumar}. 

The current-voltage curve in Figure 3c does not allow to distinguish respective reduction and oxidation peaks but rather shows an open rectangular behavior without any peaks. This at the first glance indicates double-layer-like behaviour. Distinct reaction peaks may, however, also no longer be visible in the case that several reduction steps overlap each other \cite{Supercaps}. The distinction between double layer charging and faradaic reactions can in this case be made by a kinetic analysis. Whereas double layer charging is expected to be an extremely fast process, the rate of a reduction process is limited by the diffusion of ions. Respective scan rate dependent measurements for our experiments are presented in the inset in Figure 3c. They reveal that indeed the voltammetric charge is proportional to the square root of rate, which indicates diffusion control. Together with the absence of distinct reaction peaks this behaviour is known from pseudocapacitances in metal oxides. There, the modification of the oxidation state is coupled to rate determining ion exchange processes \cite{Ardizzone}. This situation clearly differs from double layer charging which should be almost rate independent. Hence, the measured rate dependence excludes pure band filling without chemical change and instead points to redox reactions at the surface.

The potential-induced modification of the surface iron oxide layer significantly affects the overall magnetization because ultrathin FePt films are used. The 2 nm films considered here consist of about five FePt atomic double-layers. Thus, reversible oxidation of a single Fe layer may considerably reduce the magnetization of the film. This can explain the observed 5\% reduction of the saturation magnetization, even if not the whole surface would be Fe-terminated. Since FePt is a strong ferromagnet with fully occupied majority spin band, the spin-dominated magnetization of the remaining, unoxidized atomic layers is expected to be merely unaffected by the modification of the surface. On the other hand, the magnetocrystalline anisotropy of ultrathin films is a quantity which is very sensitive to structural or chemical modifications. For example, the magnetocrystalline anisotropy of Fe-terminated FePt films with 9 monolayers thickness is almost 50\% smaller than the magnetocrystalline anisotropy of films with the same thickness but Pt-termination \cite{Zhang phd}. It is therefore conceivable that chemical modification of a single atomic layer may alter the magnetocrystalline anisotropy of a 2-nm magnetic film by as much as 25\%, as observed here. Such a sensitivity has two microscopic reasons: First, orbital magnetic properties like the magnetocrystalline anisotropy depend more strongly on the band filling than the spin moment, see e.g. reference \cite{Steinbeck} and references therein; second, quantum oscillations additionally modify the magnetocrystalline anisotropy in the case of thin films \cite{Zhang}.

For thicker composite films, a potential-induced change of anisotropy based on electrochemical modification of the surface layer can be discussed within the concept of exchange coupling between hard and soft phases. In general, exchange coupling of hard and soft magnetic phases significantly reduces the effective anisotropy in comparison to the isolated hard magnetic phase. This has been exploited irreversibly already for hard magnetic FePt nanocomposites coupled to soft magnetic Fe \cite{Breitling} or iron oxide phases \cite{Zeng}. For complete coupling, the extension of the soft magnetic layer must not exceed the domain wall width of the hard magnetic phase \cite{Schrefl}, which is about 4-5 nm for FePt \cite{Hinzke}. For thin composite FePt(001)/Fe-O films with layer thickness in this dimension, the soft magnetic magnetite or Fe layer formed at negative charges can thus lead to a decreased perpendicular anisotropy due to exchange coupling. This means that an external potential can be used to switch between a magnetic exchange coupled compound (e.g. FePt/Fe) with low anisotropy and a FePt(001)/Fe-O film with non-magnetic surface oxide layer and higher anisotropy.

\begin{figure}
\includegraphics[width=0.5\columnwidth]{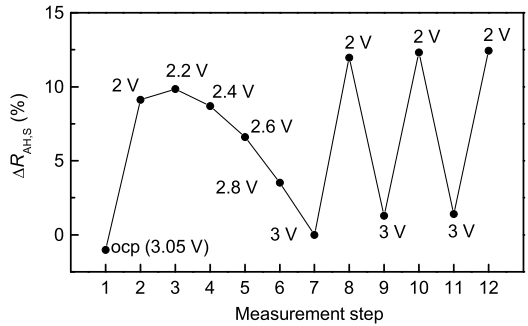}
\caption{Increased electric field effect for a composite FePt/Fe-O sample with additional Fe-O layer originating from oxidizing a 0.4 nm Fe layer ontop of a 2 nm FePt thin film. The change of $R_{\rm AH,S}$ as a measure for the saturation magnetization is plotted for several measurement steps involving voltages of and inbetween 2 and 3 V in 0.1 M LiClO$_4$ in DMC/EC (1:1). }
\label{Fig6}
\end{figure}

Up to this point, we discussed changes in the magnetic properties resulting from natural oxidation of the FePt film. Natural oxidation strongly depends on the time and kind of storage, which was not controlled in detail for our samples. Accordingly, variations of oxide layer thickness and properties may occur and may result in different magnitudes of the effect for the individual samples (\cite{Leistnerelacta}, Figure 3 and 4). Better control and an enhanced effect is expected for adding an Fe-O layer intentionally. An artificial FePt/Fe-O composite was prepared by subsequent PLD of 0.4 nm Fe ontop of a 2 nm FePt film, followed by oxidation in ambient condtions. As the deposition of the Fe layer destroyed the order of the underlying FePt and thus inhibited magnetocrystalline anisotropy, only the change in magnetization during charging is evaluated for this case. Figure 7 shows that indeed a substantially increased reversible change in magnetization of about 13\% is achieved with the thicker oxide layer. This approves the key role of the Fe-O layer and promises even larger effects when the layer structure is further optimized and better FePt order is achieved. Besides the oxide layer also the hard magnetic layer can be tuned. On step in this direction has been carried out by Reichel et al. \cite{Reichel} who used CoPt/Co-O instead of FePt/Fe-O. They reported large potential-induced coercivity variations that also originated from redox reactions in the oxide layer. Both examples show that the overall potential dependent magnetic behaviour of FePt(CoPt)/oxide composites can be widely tuned by optimizing one or both of the composite layers.

At this point, a comparison with the study of Weisheit et al. \cite{Weisheit} is attempted that dealt with a similar system (ultrathin FePt films) and setup (charging in an electrolyte). Whereas Weisheit et al. \cite{Weisheit} measure 5\% change in coercivity and 3\% change in Kerr rotation for a potential change of -0.6 V, we get up to almost 30\% change in anisotropy and 5\% change in $R_{\rm AH,S}$ for a potential change of -1 V. Besides the increased potential window, one reason for the larger effects in our case certainly is the establishment of a critical situation by a tuned moderate magnetocrystalline anisotropy. In contrast to the highly ordered films used by Weisheit et al. \cite{Weisheit}, small changes will then have a larger effect on the overall magnetization behaviour. In contrast to the reversible reduction/oxidation applied in our study, Weisheit et al. \cite{Weisheit} explain the reduced coercivity by a decrease of magnetocrystalline anisotropy by electronic (double layer) charging. This explanation was subsequently confirmed by electron theory to yield the correct sign and magnitude \cite{Zhang}. Since they expected FePt to be Pt terminated and thus inert even in ambient conditions, Weisheit et al. \cite{Weisheit} did not study their films with regard to surface oxidation. It is thus not possible to exclude oxide-related effects for their films. The present results indicate, however, that the role of surface oxide should always be analyzed when direct or indirect potential dependent changes of magnetization (which includes the amplitude of Kerr rotation) are obtained. This goes along with several more recent studies reporting on oxygen-related mechanisms for electric field control of properties: Redox reactions in Fe-oxides are discussed for the magnetization changes in Fe-O/Pt nanocomposites \cite{Traussnig}; FeO interlayers are made responsible for a change of anisotropy in Fe/MgO \cite{Nakamura}; the creation of oxygen vacancies is used to explain the suppression of the metal-insulator transition in VO$_2$ \cite{Jeong}. We suggest to (re-)investigate the relevance of oxide-related mechanisms also on previous experiments on electric field control of magnetic properties \cite{Weisheit,Gosh,Zhernenkov}, especially since sometimes experiment and explanation by electron theory differ by sign \cite{Gosh,Zhernenkov}.  

\section{CONCLUSION}
To conclude, we demonstrated that electric field control of magnetism is possible on the basis of reversible reduction/oxidation in metal oxides. For this goal, ultrathin composite films are of benefit in two ways: (1) they allow adjusting a critical point between shape and magnetocrystalline anisotropy and (2) their properties can essentially be controlled by electrochemical modification of surface atomic layers. By the charge-induced change between surface compositions with different magnetic properties at a critical point, a variation of anisotropy by 25\% is obtained, compared to 5\% by assumed electrical charging of highly ordered FePt thin films \cite{Weisheit}. Our electrochemical analysis reveals that E-field induced variations of saturation magnetization can be attributed to oxidation/reduction processes. Similar redox processes are involved during resistive switching of thin oxide layers, a novel route for nonvolantile storage \cite{Waser}. Our approach is versatile and can be extended to various metal/oxide combinations and  also thicker exchange coupled composites which allows to tailor magnetoelectric properties in a broad manner. 

We acknowledge H. Schl\"orb for discussion. This work is funded by DFG by project LE2558/1-1.


\end{document}